\documentclass[prl,floatfix,twocolumn,showpacs,superscriptaddress]{revtex4}
\pdfoutput=1
\usepackage{graphicx}
\usepackage{amssymb}
\usepackage{amsmath,amsfonts}
\usepackage{epstopdf}

\newcommand {\ket}[1] {|#1 \rangle}

\newcommand {\ketbra}[2] {| #1 \rangle \langle #2 |}

\newcommand {\av}[1] {\langle #1 \rangle}

\begin{document}

\title{Dynamical quantum phase transitions in the dissipative Lipkin-Meshkov-Glick model with proposed realization in optical cavity QED}

\author{S. Morrison}
\affiliation{Institute for Theoretical Physics, University of Innsbruck, Innsbruck A-6020, Austria}
\affiliation{Institute for Quantum Optics and Quantum Information of the Austrian Academy of
Sciences, A-6020 Innsbruck, Austria}

\author{A.~S. Parkins}
\affiliation{Department of Physics, University of Auckland, Private Bag 92019, Auckland, New Zealand}

\date{\today}

\begin{abstract}
We present an optical cavity QED configuration that is described by
a dissipative version of the Lipkin-Meshkov-Glick model of an
infinitely coordinated spin system. This open quantum system
exhibits both first- and second-order non-equilibrium quantum phase
transitions as a single, effective field parameter is varied. Light
emitted from the cavity offers measurable signatures of the critical
behavior, including that of the spin-spin entanglement.

\end{abstract}


\maketitle

%%%%%%%%%%%%%%%%%%%%%%%%%%%%%%%%%%%%%%%%%%%%%%%
%%%%%%%%%%%%%%%%%%%%%%%%%%%%%%%%%%%%%%%%%%%%%%%

Remarkable advances with trapped, ultra-cold atomic gases have
opened up exciting new avenues of research into strongly interacting
many-body quantum systems \cite{Bloch05}. Exquisite control of both
motional and  electronic degrees of freedom of cold atoms can enable
one to ``tailor'' atom-atom interactions and thereby implement a
variety of systems that exhibit, in particular, quantum critical
phenomena, i.e., transitions between distinct quantum phases, driven
by quantum fluctuations, in response to variations of an effective
field or interaction strength around some critical value.

Recently, important insights into such transitions have been
obtained from theoretical studies of the quantum entanglement
properties of critical spin systems (see, e.g.,
\cite{Osterloh02,Osborne02,GVidal03,EntLMGFirstOrder,EntLMGSecondOrder,EntLMGEntropy,Fleischhauer05,Lambert04,Reslen05}).
Bipartite entanglement measures characterizing entanglement between
a pair of spins (e.g., the concurrence) or between two blocks of
spins (e.g., the entanglement entropy) can display marked critical
behavior and scaling at quantum critical points. In this context, a
simple but very useful example is the Lipkin-Meshkov-Glick (LMG)
model \cite{originalLMG123}, which is described by the Hamiltonian
\begin{equation} \label{LMGHamiltonian}
H_\textrm{LMG} =  -2 h J_z  - (2\lambda/N) (J_x^2 +\gamma J_y^2),
\end{equation}
where $\{ J_x,J_y,J_z\}$ are collective angular momentum operators
for $N$ spin-1/2 particles, $h$ and $\lambda$ are effective magnetic
field and spin-spin interaction strengths, respectively, and
$\gamma\in[-1,1]$ is an anisotropy parameter. This system, in which
each spin interacts identically with every other spin, exhibits
critical behavior at zero temperature; in particular, either first-
or second-order equilibrium quantum phase transitions may occur,
depending on the choice of $\lambda$ and $\gamma$, as the ratio
$h/\lambda$ is varied across a critical value
\cite{EntLMGSecondOrder}. Notably, the second-order transition
involves a change from a unique ground state (normal phase) to a
pair of macroscopically displaced degenerate ground states (broken
phase). Entanglement in the system displays the above-mentioned
critical behavior, reaching, in particular, a pronounced maximum at
the critical point
\cite{EntLMGFirstOrder,EntLMGSecondOrder,EntLMGEntropy}.

Given these interesting and topical features of the LMG model, it
follows that the physical realization of a system described by such
a model would provide a valuable test bed for studies of quantum
critical phenomena and entanglement. Here we propose an open-system
(i.e., dissipative) version of the LMG model based on the collective
interaction of an ensemble of atoms with laser fields and field
modes of a high-finesse optical resonator. In the spirit of a recent
proposal for realizing the Dicke model \cite{Dimer07}, our scheme
employs Raman transitions between a pair of atomic ground states and
the relevant energy scales (e.g., $h$, $\lambda$) are set by light
shifts of the atomic levels and Raman transition rates and
detunings.

The open nature of this system, a consequence of the external
driving fields and cavity mode losses, introduces a number of
important differences from, and, arguably, advantages over, the
closed, Hamiltonian LMG system: (i) thermal equilibrium phase
transitions are replaced by dynamical, non-equilibrium phase
transitions, (ii) the cavity output fields offer quantitative
measures of properties of the collective-spin system, including
entanglement, in the critical regime, and (iii) it is possible to
observe both first- and second-order quantum phase transitions as a
{\em single effective field parameter, $h$, is varied}.

%%%%%%%%%%%%%%%%%%%%%%%%%%%%%%%%%%%%%%%%%%%%%%%

\begin{figure}[h!t]
\centerline{\includegraphics[width=8.6cm]{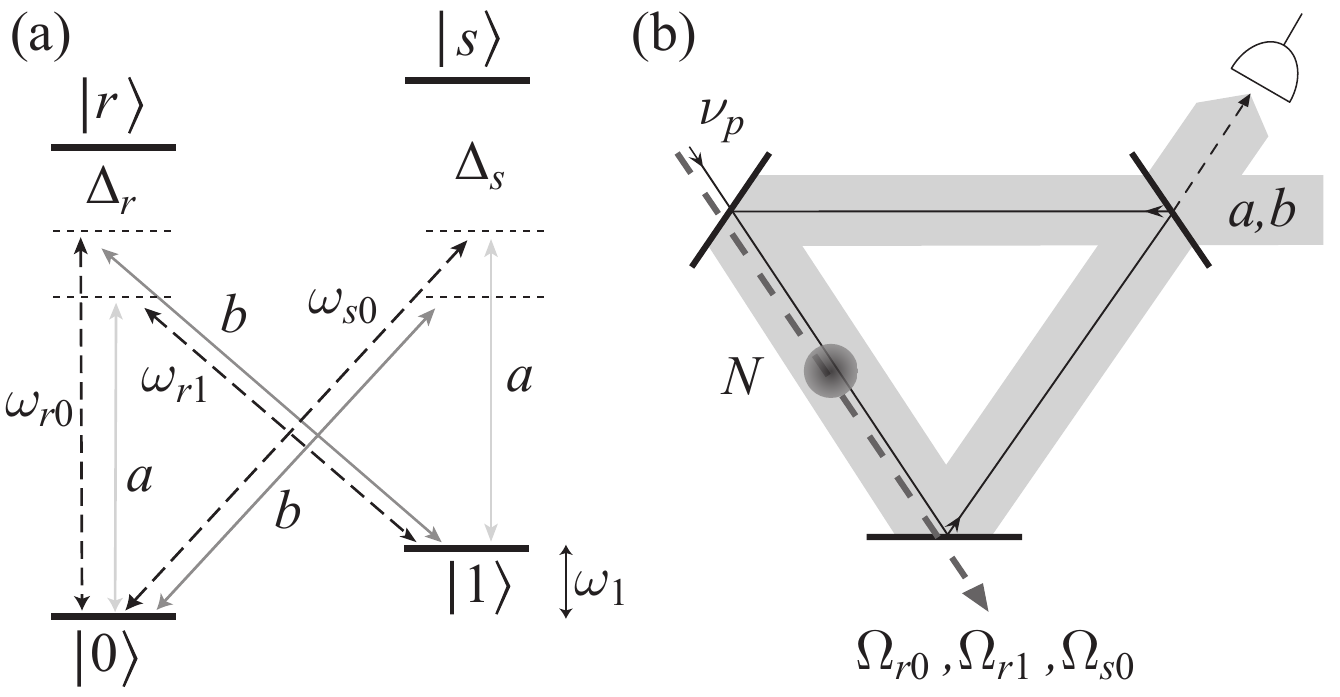}}
\caption{(a) Atomic level and excitation scheme. (b) Potential
ring-cavity setup. The laser fields (dashed lines) are at
frequencies that are not supported by the resonator, but can be
injected through one of the resonator mirrors so as to be
co-propagating with the cavity fields through the ensemble.}
\label{fig:atomic_level_scheme}
\end{figure}

We consider $N$ atoms coupled via electric dipole transitions to
three laser fields and to a pair of independent (e.g.,
orthogonally-polarized) optical cavity modes. The atomic level and
excitation scheme is shown in Fig.~\ref{fig:atomic_level_scheme},
together with a possible ring-cavity setup. At the location of the
atoms, the cavity and laser fields are copropagating traveling
waves, with sufficiently broad beam waists so as to ensure
homogeneous atom-field couplings. These fields combine to drive
Raman transitions between two stable electronic ground states of the
atoms, $\ket{0}$ and $\ket{1}$ (energies $\omega_0=0$ and
$\omega_1$, respectively, with $\hbar=1$) via the excited atomic
states $\ket{r}$ and $\ket{s}$ (energies $\omega_r$ and $\omega_s$).
The laser fields have optical frequencies $\omega_{r0}$,
$\omega_{s0}$, and $\omega_{r1}$, and couple to the atomic
transitions with Rabi frequencies $\Omega_{r0}$, $\Omega_{s0}$, and
$\Omega_{r1}$. Cavity field $a$, at frequency $\omega_a$, couples to
the transitions $\ket{0} \leftrightarrow \ket{r}$ and $\ket{1}
\leftrightarrow \ket{s}$ with strengths $g_{r0}$ and $g_{s1}$,
respectively, while cavity field $b$, at frequency $\omega_b$,
couples to the transitions $\ket{0} \leftrightarrow \ket{s}$ and
$\ket{1} \leftrightarrow \ket{r}$ with strengths $g_{s0}$ and
$g_{r1}$, respectively. As drawn in
Fig.~\ref{fig:atomic_level_scheme}, the level scheme would apply,
e.g., to ${}^{6}{\rm Li}$, with the ground magnetic substates
$|F=1/2,m=\pm1/2\rangle$ as $|0\rangle$ and $|1\rangle$ and a
magnetic field perpendicular to the cavity axis to provide a
splitting $\omega_1$ between these states. Modes $a$ and $b$ would
be orthogonal, linearly-polarized cavity modes, with mode $a$
polarized along the direction of the magnetic field.

The atom-light detunings $\Delta_r=\omega_r-\omega_{r0}$ and
$\Delta_s=\omega_s-\omega_{s0}$ are taken to be much larger than any
dipole coupling strengths, atomic linewidths, or cavity loss rates.
This enables us to adiabatically eliminate the states $\ket{r}$ and
$\ket{s}$ from the dynamics and neglect the effects of atomic
spontaneous emission. Additionally, as depicted in
Fig.~\ref{fig:atomic_level_scheme}, we assume that only three
distinct Raman transitions are of significance (i.e., resonant or
roughly resonant); i.e., we retain only those Raman processes that
cause a change in the electronic state of the atoms
($\ket{0}\rightarrow\ket{1}$ or $\ket{1}\rightarrow\ket{0}$) and
also involve transfer of a photon from a laser field to a cavity
mode or vice-versa. All other possible Raman processes are assumed
to be far off-resonant and therefore negligible. Finally, taking the
wave numbers of the laser and cavity fields to be essentially equal,
and introducing the collective spin operators $J_z = \frac{1}{2}
\sum_{j=1}^N \left(\ketbra{1_j}{1_j} - \ketbra{0_j}{0_j}\right)$,
$J_+ = \sum_{j=1}^N \ketbra{1_j}{0_j}$, and $J_- = (J_+)^\dagger$,
we can derive a master equation for the cavity modes and
ground-state atoms in the form
\begin{eqnarray}
\dot{\rho}_{\rm g} = -i[H_{\rm g},\rho_{\rm g}] +  \kappa_a D[a]\rho_{\rm g} + \kappa_b D[b]\rho_{\rm g} ,
\label{eq:master_eqn_atom_cavity}
\end{eqnarray}
where $D[A]\rho = 2 A \rho A^\dagger - A^\dagger A \rho - \rho A^\dagger A$, $\kappa_{a,b}$ are the
cavity field decay rates, and (omitting constant energy terms)
\begin{eqnarray}
H_{\rm g} &= &\omega_0 J_z +\delta_a a^\dagger a + \delta_b b^\dagger b + 2\delta_a^- J_z a^\dagger a + 2\delta_b^- J_z b^\dagger b \nonumber \\
&&  +  \frac{\lambda_a}{\sqrt{N}} J_x (a+a^\dagger) + \frac{\lambda_b}{\sqrt{N}} (J_- b+J_+ b^\dagger), \label{eq:Heff}
\end{eqnarray}
with $J_x=(J_++J_-)/2$ and
\begin{subequations}
\begin{eqnarray}
\omega_0 & = & \frac{|\Omega_{r1}|^2}{4\Delta_r} -
\frac{|\Omega_{r0}|^2}{4\Delta_r} -\frac{|\Omega_{s0}|^2}{4\Delta_s}
+ \omega_1 - \omega_1', \label{eq:def(a)}
\\
\delta_a &=& \omega_a + \omega_1' - \omega_{s0} + N\delta_a^+ ,
\\
\delta_b &=& \omega_b + \omega_1' - \omega_{r0} + N\delta_b^+ ,
\\
\delta_a^\pm & = &\frac{|g_{s1}|^2}{2\Delta_s} \pm \frac{|g_{r0}|^2}{2\Delta_r}, ~~
\delta_b^\pm =\frac{|g_{r1}|^2}{2\Delta_r} \pm \frac{|g_{s0}|^2}{2\Delta_s},
\\
\lambda_a & = & \frac{\sqrt{N}\Omega_{r1}^*g_{r0}}{\Delta_r} = \frac{\sqrt{N}\Omega_{s0}^*g_{s1}}{\Delta_s}, %\label{eq:def(d)}
%\\
~\lambda_b = \frac{\sqrt{N}\Omega_{r0}^*g_{r1}}{2\Delta_r}, ~~~~ \label{eq:def(e)}
\end{eqnarray}
\end{subequations}
where $\omega_1' =(\omega_{s0}-\omega_{r1})/2\simeq\omega_1$, and we
have assumed the two Raman transitions involving mode $a$ to occur
at the same rate $\lambda_a$.

%%%%%%%%%%%%%%%%%%%%%%%%%%%%%%%%%%%%%%%%%%%%%%%

We now assume $(\kappa_i^2+\delta_i^2)^{1/2}\gg
\lambda_a,\lambda_b,\omega_0$. In this limit, the cavity modes are
only weakly or virtually excited and may also be adiabatically
eliminated to yield the following master equation for the reduced
density operator, $\rho$, of the collective atomic system alone:
\begin{eqnarray}
\dot{\rho} & = & -i[H_{LMG}^{\gamma=0},\rho] + \frac{\Gamma_a}{N}
D[2J_x]\rho + \frac{\Gamma_b}{N} D[J_+]\rho ,
\label{eq:master_equation_gamma0_lmg_model}
\end{eqnarray} with
$h=-\omega_0/2$, $\lambda =
2\lambda_a^2\delta_a/(\kappa_a^2+\delta_a^2)$, and $\Gamma_i =
\lambda_i^2\kappa_i/(\kappa_i^2+\delta_i^2)$ ($i=a,b$). Note that in
deriving (\ref{eq:master_equation_gamma0_lmg_model}) we have also
assumed that $\kappa_b\gg\delta_b\simeq 0$. If we then take
$\delta_a\gg\kappa_a$ and $\Gamma_a\ll\Gamma_b$, then the role
played by each cavity mode in relation to the atomic system is quite
distinct. Specifically, mode $a$ mediates the collective spin-spin
interaction (of strength $\lambda \simeq \lambda_a^2/\delta_a$)
associated with the Hamiltonian dynamics, whilst mode $b$ mediates
the collective atomic decay (with rate
$\Gamma_b\simeq\lambda_b^2/\kappa_b$).

%%%%%%%%%%%%%%%%%%%%%%%%%%%%%%%%%%%%%%%%%%%%%%%

\begin{figure} [h!t]
\centerline{\includegraphics[width=8.4cm]{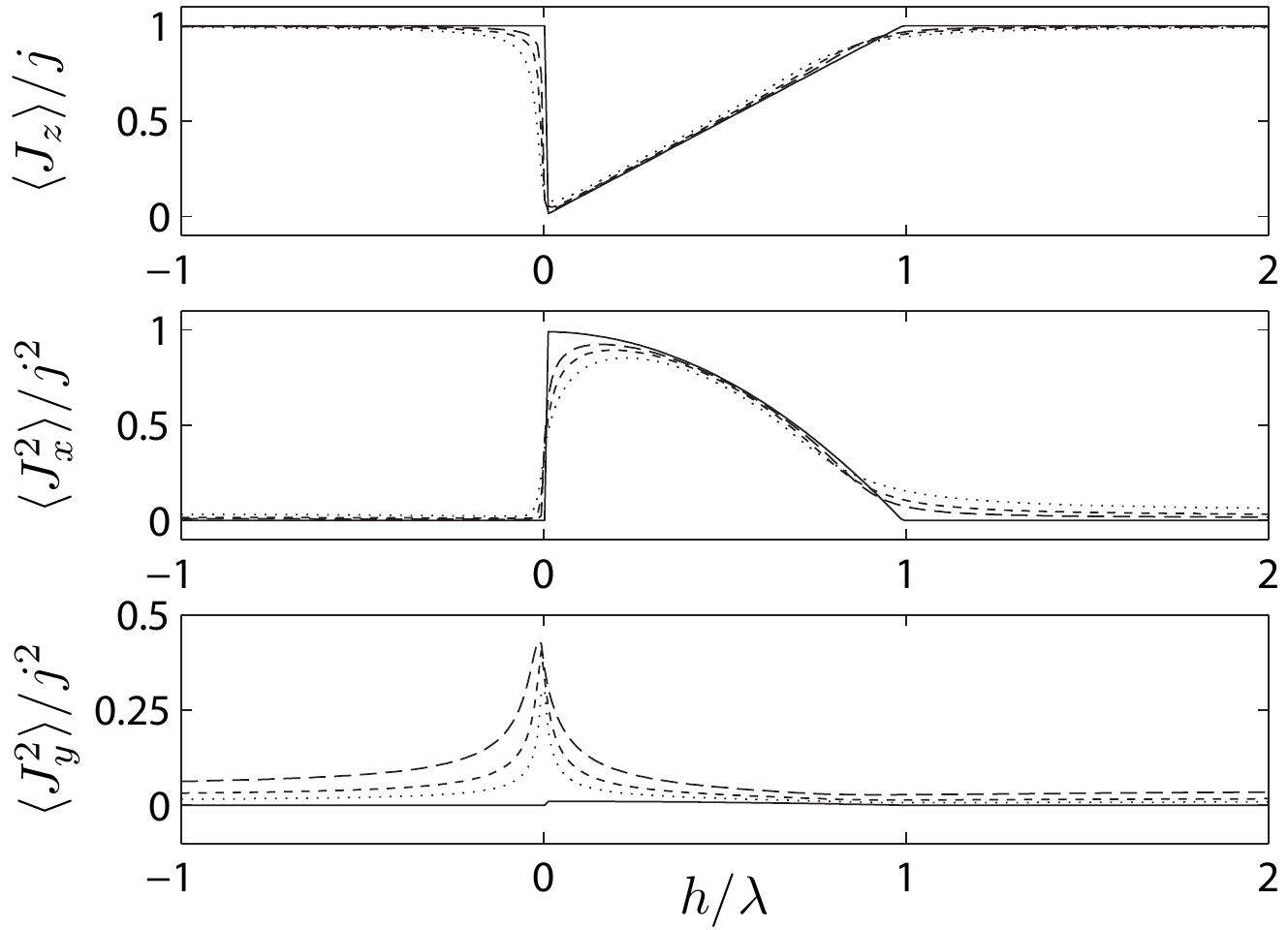}}
\caption{Semiclassical (solid line) and finite-$N$ steady-state
inversion and second-order moments for $\Gamma_a/\lambda = 0.01$,
$\Gamma_b/\lambda=0.2$, and $N = 25$ (dotted), $50$ (short dash),
$100$ (long dash).} \label{fig:semicl}
\end{figure}

The equations of motion for the moments
$\{\av{J_x},\av{J_y},\av{J_z}\}$, derived from
(\ref{eq:master_equation_gamma0_lmg_model}), do not form a closed
set. However, factorizing the means of operator products and taking
the limit $N\rightarrow\infty$ (i.e., neglecting quantum
fluctuations), we obtain a closed set of semiclassical equations,
\begin{subequations}
\begin{eqnarray}
\dot{X} & = & 2h Y - \Gamma_b Z X,  \label{eq:semicl(a)}\\
\dot{Y} & = & -2h X + 2\lambda Z X - \Gamma_b Z Y, \label{eq:semicl(b)}\\
\dot{Z} & = & -2\lambda X Y + \Gamma_b(X^2 +Y^2), \label{eq:semicl(c)}
\end{eqnarray}
\end{subequations}
where $(X,Y,Z)\equiv(\av{J_x},\av{J_y},\av{J_z})/j$ with $j=N/2$,
and $X^2 + Y^2 + Z^2 = 1$ (conservation of angular momentum). The
stable steady-state solutions of (6) exhibit bifurcations at {\em
two} critical effective field strengths, $h_\pm^{\rm c} =
[\lambda\pm (\lambda^2-\Gamma_b^2)^{1/2}]/2$ (we assume $\lambda
>0,\Gamma_b$). In particular, for $h<h_-^{\rm c}$ and $h>h_+^{\rm
c}$ the stable steady-state solutions are $\{X_{\rm ss}=Y_{\rm
ss}=0,~Z_{\rm ss}=1\}$, whereas for $h_-^{\rm c}<h<h_+^{\rm c}$ one
finds
\begin{eqnarray}
X_\textrm{ss} = \pm \sqrt{\frac{\Lambda^2-4h^2}{2\lambda\Lambda}}, ~~
Y_\textrm{ss} = \frac{\Gamma_b}{\Lambda} X_\textrm{ss} , ~~
Z_\textrm{ss} = \frac{2h}{\Lambda},
\end{eqnarray}
where $\Lambda = \lambda + (\lambda^2-\Gamma_b^2)^{1/2}$. Note that
at both (supercritical pitchfork) bifurcations a detailed stability
analysis \cite{Morrison07} shows that a unique steady state becomes
unstable and two new stable steady states emerge. These
semiclassical solutions, together with numerical solutions of the
finite-$N$ master equation
(\ref{eq:master_equation_gamma0_lmg_model}), are plotted in
Fig.~\ref{fig:semicl} as a function of $h/\lambda$ (note that
$\av{J_x}=\av{J_y}=0$ for the finite-$N$ calculations). The plots
indicate both a first- and second-order phase transition {\em as a
single parameter, h, is varied}. The first-order (second-order)
transition, at $h=h_-^{\rm c}$ ($h=h_+^{\rm c}$), involves a
discontinuous (continuous) bifurcation in $X_{\rm ss}$ and
associated behavior in $Z_{\rm ss}$. Note that in the purely
Hamiltonian system second-order transitions occur at $\pm h_+^c$,
but the first-order transition has no counterpart (for $\lambda>0$)
and arises here due to a dissipative instability. The behavior we
observe bears some relation to critical points found in cooperative
resonance fluorescence (see, e.g., \cite{Bonifacio76Drummond78}).

%%%%%%%%%%%%%%%%%%%%%%%%%%%%%%%%%%%%%%%%%%%%%%%

In the large-$N$ limit, quantum fluctuations can be included in the
analysis as a first-order correction using a large-$N$ expansion of
the Holstein-Primakoff (HP) representation of angular momentum
operators \cite{Holstein40}. Applied in a coordinate system where
the mean Bloch vector points along the positive $z$-axis, this takes
the form $J_z=N/2 -c^\dagger c\simeq N/2$ and $J_+=(N-c^\dagger
c)^{1/2}c\simeq \sqrt{N}\, c$, where $c$ ($c^\dagger$) is a bosonic
annihilation (creation) operator. This linearization about the mean
field state leads to a master equation of the general form
\begin{eqnarray} \label{eq:linearised_master_equation}
\dot{\rho} & = & -i[H_{\textrm{HP}},\rho] + \Gamma_{+} D[c^\dagger]\rho + \Gamma_{-} D[c]\rho \nonumber \\
&+& \left\{ \Upsilon \left( 2c\rho c -c^2\rho - \rho c^2 \right) + \textrm{H.c.} \right\} ,
\label{eq:damping_spin_louvillian}
\end{eqnarray}
where $H_{\textrm{HP}}$ is a quadratic in $\{ c,c^\dagger\}$ and the
coefficients are functions of $\{ h,\lambda,\Gamma_a,\Gamma_b\}$
\cite{Morrison07}. Eq.~(\ref{eq:damping_spin_louvillian}) yields
coupled, linear equations of motion for $\av{c}$ and
$\av{c^\dagger}$, the eigenvalues of which display a sequence of
bifurcations in both their real and imaginary parts as $h$ is
varied. The phase transitions are marked by the real part of one
eigenvalue going to zero (i.e., critical slowing down) at
$h=h_\pm^{\rm c}$.

%%%%%%%%%%%%%%%%%%%%%%%%%%%%%%%%%%%%%%%%%%%%%%%

To examine this structure and dynamics, we consider the transmission
of a (weak) probe laser field through the medium as a function of
the probe frequency, i.e., we examine the frequency response of the
system. A schematic of such a measurement setup is shown in
Fig.~\ref{fig:atomic_level_scheme}(b). To compute the transmission
spectrum we retain the two cavity modes in our model (i.e., we start
from (\ref{eq:master_eqn_atom_cavity})), but again perform a
linearization ($N\gg 1$) about the mean-field state. We consider the
case in which the probe laser drives mode $b$, and the transmission
spectrum $T_{p}(\nu_{p})$ is defined as the coherent intensity, at
probe frequency $\nu_p$ (in the rotating frame), in the output field
from mode $b$.

 \begin{figure}[h!t]
\includegraphics[width=8.6cm]{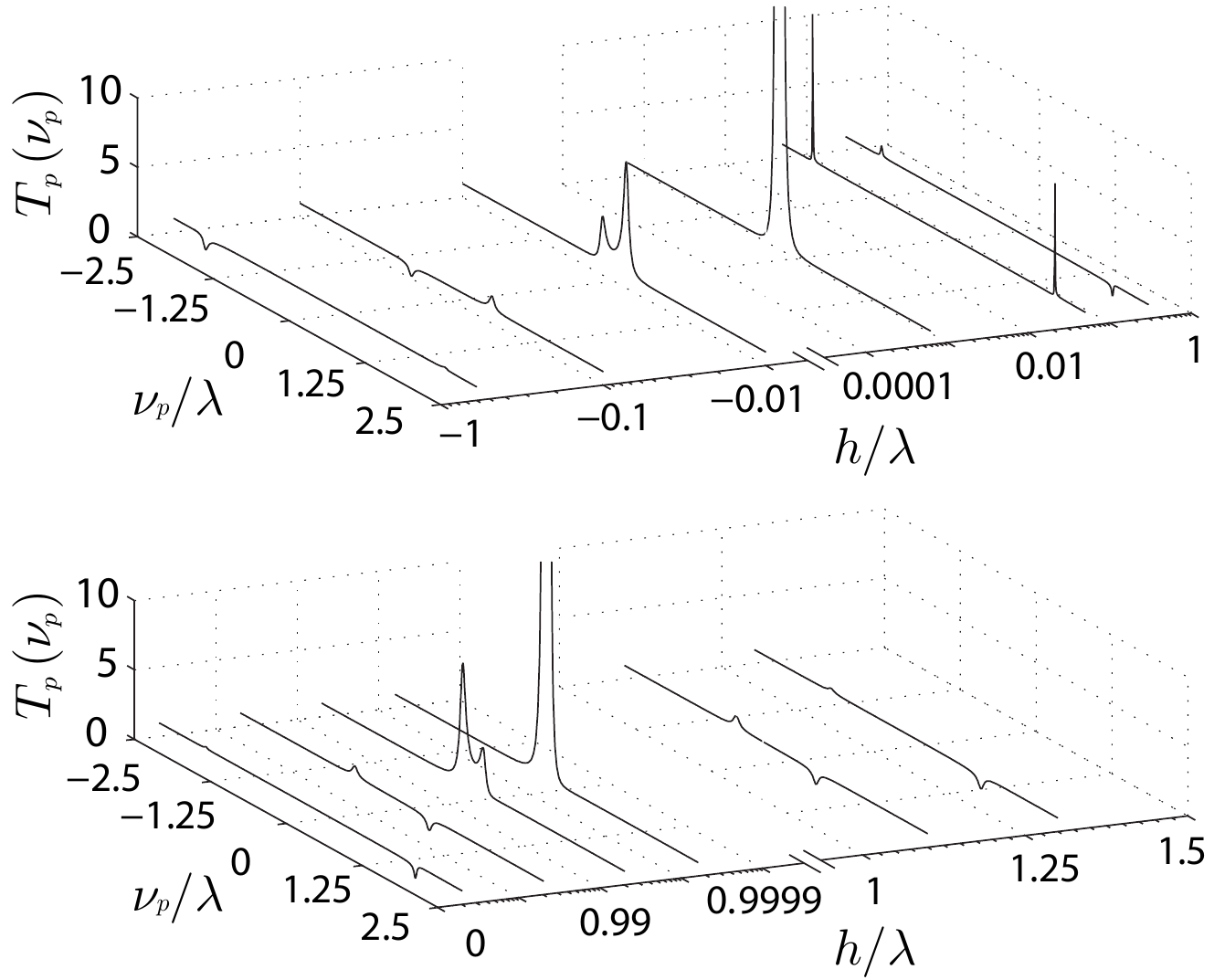}
\caption{Transmission spectra  in the linearised regime, for
$h/\lambda =\{-0.6,-0.1,-0.01,h_-^{\rm c}/\lambda,0.05,0.3\}$ (top),
and $\{0.5,0.95,0.995,h_+^{\rm c}/\lambda,1.1,1.3\}$ (bottom), with
microscopic parameters
$\kappa_a/\delta_a=0.02,\,\lambda_b/\lambda_a=0.32,\,\kappa_b/\delta_a=1,\,\delta_b=0$,
giving $\Gamma_a/\lambda=0.01,\,\Gamma_b/\lambda=0.05$. We set
$\delta_{a,b}^-=0$.} \label{fig:trans_spect_1}
\end{figure}

In Fig. \ref{fig:trans_spect_1} we plot $T_{p}(\nu_{p})$ (normalized
by the maximum empty-cavity transmission) for a series of values of
$h$ around $h_\pm^{\rm c}$. For the chosen parameters, the spectra
consist of sharp ``atomic'' resonances superimposed on a much
broader cavity mode resonance (i.e., $\kappa_b\gg\Gamma_b$). The
locations and widths of the atomic resonances are determined by the
imaginary and real parts of the above-mentioned eigenvalues,
respectively. For $|h/\lambda|>1$, the main atomic feature is a dip
of width $2\Gamma_b$ at $\nu\simeq 2h$, corresponding to a
cavity-mediated, collective spontaneous emission resonance. For
$|h/\lambda|<1$, spin-spin interactions play a more significant role
and a pair of resonances at opposite frequencies feature in the
spectrum. As $h\rightarrow h_+^{\rm c}$ both from above and below
these two features merge continuously into a single peak, centered
at $\nu_p=0$, which ultimately diverges at $h=h_+^{\rm c}$ in a
pronounced signature of the second-order phase transition. The same
merging and divergence is seen for the first-order transition, but
only as $h\rightarrow h_-^{\rm c}$ from below. For $h$ very small
(but $>h_-^{\rm c}$), the spectrum consists of two sharp peaks of
width $\sim\Gamma_bh/\lambda$ at $\nu_p\simeq \pm2\lambda$. The
transition is signaled by a discontinuous jump from this two-peaked
spectrum to a single divergent peak at $\nu_p=0$.

%%%%%%%%%%%%%%%%%%%%%%%%%%%%%%%%%%%%%%%%%%%%%%%

\begin{figure}[h!t]
\includegraphics[width=8.6cm]{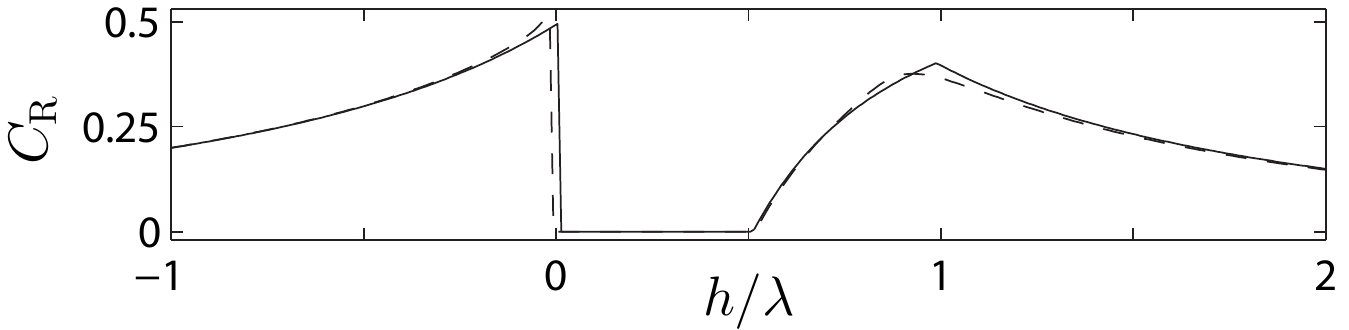}
\caption{Maximum entanglement $C_\textrm{R}$ computed from the
linearized HP ($N\rightarrow\infty$) model (solid) and from
(\ref{eq:master_equation_gamma0_lmg_model}) for $N=100$ (dashed),
with $\Gamma_a/\lambda = 0.01$ and $\Gamma_b/\lambda=0.2$.}
\label{fig:fNcss}
\end{figure}

To analyze the entanglement properties of the system, we adopt a
criterion for bipartite entanglement in collective spin systems
which, for symmetric states, is both necessary and sufficient, and
reads \cite{EntanglementCriteria}
\begin{equation}
C_{\varphi} \equiv 1-(4/N)\av{\Delta J_{\varphi}^2}-(4/N^2)\av{J_{\varphi}}^2 > 0, \label{eq:finite_N_ent_criteria}
\end{equation}
where $J_\varphi = \sin(\varphi)J_x + \cos(\varphi) J_y $. Here, we
present numerical results for $C_{\rm R}\equiv\max_\varphi
C_\varphi$ ($\geq 0$), which, in fact, equals the rescaled
concurrence $(N-1)C$, where $C$ is the two-spin concurrence. In
Fig.~\ref{fig:fNcss} we plot the steady state value of
$C_\textrm{R}$ versus $h/\lambda$, computed from the linearized HP
model and numerically from
(\ref{eq:master_equation_gamma0_lmg_model}) for finite $N$. Both
transitions are characterized by a sharp peak in the entanglement at
the critical point
\cite{EntLMGFirstOrder,EntLMGSecondOrder,EntLMGEntropy}, but they
are distinguished by a discontinuity in $C_{\rm R}$ at $h=h_-^{\rm
c}$ (for $N\rightarrow\infty$) as opposed to a discontinuity in
$\partial C_{\rm R}/\partial h$ at $h=h_+^{\rm c}$
\cite{maxC_varphi}. The peaking of $C_{\rm R}$ at the critical
points agrees with the conjecture of a general association between
semiclassical bifurcations and maximal entanglement in dissipative,
non-equilibrium many-body systems \cite{Schneider02}. In the region
where $C_{\rm R}=0$ the state approaches a mixture of maximally
polarised states possessing large fluctuations (see Fig.
\ref{fig:semicl}). Note that, in the adiabatic regime considered,
the cavity field operator $b(t) \propto J_+(t)$, and so collective
spin correlations (and hence $C_{\rm R}$) can be deduced from
moments of the cavity output field, which may be measured by
broadband homodyne detection.

%%%%%%%%%%%%%%%%%%%%%%%%%%%%%%%%%%%%%%%%%%%%%%%

For an experimental realization, we have already mentioned
${}^{6}{\rm Li}$ in a ring-cavity setup. A suitable system can also
be designed using the ground states $|F=1,m=\pm1\rangle$ of
${}^{87}{\rm Rb}$ and linearly-polarized cavity modes
\cite{Morrison07}. For specific parameter values, we consider recent
experiments with cold atoms inside a high-finesse optical ring
cavity \cite{vonCube06}, i.e., we take $g_{ij}\simeq 2\pi\cdot
100~\textrm{kHz}$ and $\kappa_a\simeq 2\pi\cdot 25~\textrm{kHz}$.
For $N\simeq 10^6$ atoms and a characteristic ratio
$\Omega_{ij}/\Delta_i\simeq 0.0025$, we have $\lambda_{a}\simeq
2\pi\cdot 250~\textrm{kHz}$. With a Raman detuning $\delta_{a}\simeq
2\pi\cdot 2.5~\textrm{MHz}$, we then have $\lambda\simeq
2\lambda_{a}^2/\delta_{a} \simeq 2\pi\cdot 25~\textrm{kHz}$ and
$\Gamma_{a} \simeq \kappa_{a} (\lambda_a/\delta_a)^2 \simeq
2\pi\cdot 0.25~\textrm{kHz}$. Ground state magnetic level shifts of
tens of MHz would suffice to ensure distinct Raman channels. Mode
$b$ may be more strongly damped (i.e., the two cavity polarizations
have different finesses), e.g., $\kappa_{b}\simeq 2\pi\cdot
250~\textrm{kHz}$, and, with $\lambda_{b}\simeq 2\pi\cdot
25~\textrm{kHz}$ and $\delta_{b}\simeq 0$, we would then have
$\Gamma_{b} \simeq\lambda_{b}^2/\kappa_{b} \simeq 2\pi\cdot
2.5~\textrm{kHz}\gg\Gamma_{a}$. Finally, the rate for single-atom
spontaneous emission (neglected in our model) is estimated by
$\Gamma_{\rm at}\Omega_{ij}^2/(4\Delta_i^2) \lesssim 2\pi \cdot
0.01~\textrm{kHz} \ll \lambda,\Gamma_{b}$ for an atomic exited state
decay rate $\Gamma_{\rm at}=2\pi\cdot 6~\textrm{MHz}$.

To conclude, we have proposed a feasible cavity QED system that is
described by a dissipative LMG model and exhibits both first- and
second-order non-equilibrium quantum phase transitions as a function
of a single effective field parameter. Measurements on the cavity
output light fields provide quantitative probes of the critical
behavior. The system also offers opportunities for investigating
phase transitions in response to variation of the strength of
dissipation (i.e., $\Gamma_b$), for studying time-dependent
behavior, such as entanglement dynamics, and for preparing very
highly entangled states, which typically occur for short interaction
times \cite{Morrison07} and may in principle be ``frozen" by
switching off all optical fields.

The authors thank A. Daley and H. Carmichael for discussions and
acknowledge support from the Austrian Science Foundation and from
the Marsden Fund of the Royal Society of New Zealand.

%%%%%%%%%%%%%%%%%%%%%%%%%%%%%%%%%%%%%%%%%%%%%%%
%%%%%%%%%%%%%%%%%%%%%%%%%%%%%%%%%%%%%%%%%%%%%%%


\begin{thebibliography}{99}

\bibitem{Bloch05}
I. Bloch, Nature Phys. {\bf1}, 23 (2005).


\bibitem{Osterloh02}
A. Osterloh {\em et al}., Nature {\bf416}, 608 (2002).

\bibitem{Osborne02}
T.~J. Osborne and M.~A. Nielsen, Phys. Rev. A {\bf66}, 032110 (2002).

\bibitem{GVidal03}
G. Vidal {\em et al}., Phys. Rev. Lett. {\bf90}, 227902 (2003).

\bibitem{EntLMGFirstOrder}
%J. Vidal, G. Palacios, and R. Mosseri, Phys. Rev. A {\bf69}, 022107 (2004);
J. Vidal, R. Mosseri, and J. Dukelsky, Phys. Rev. A  {\bf69}, 054101 (2004).

\bibitem{EntLMGSecondOrder}
S. Dusuel and J. Vidal, Phys. Rev. B {\bf 71}, 224420 (2005)

\bibitem{EntLMGEntropy}
J. Latorre {\em et al}., Phys. Rev. A {\bf71}, 064101 (2005).

\bibitem{Fleischhauer05}
R.~G. Unanyan, C. Ionescu, and M. Fleischhauer,  Phys. Rev. A {\bf72}, 022326 (2005).

\bibitem{Lambert04}
N. Lambert, C. Emary, and T. Brandes, Phys. Rev. Lett. {\bf92},  073602 (2004); Phys. Rev. A {\bf71},  053804 (2005).

\bibitem{Reslen05}
J. Reslen, L. Quiroga, and N.~F. Johnson, Europhys. Lett. {\bf69}, 8 (2005).


\bibitem{originalLMG123}
H.~J. Lipkin, N. Meshkov, and A.~J. Glick, Nucl. Phys.  {\bf62}, 188 (1965).


\bibitem{Dimer07}
F. Dimer {\em et al}., Phys. Rev. A {\bf75}, 013804 (2007).


\bibitem{Bonifacio76Drummond78}
R. Bonifacio and L.~A. Lugiato, Opt. Commun. {\bf19}, 172 (1976);
P.~D. Drummond and H.~J. Carmichael, Opt. Commun. {\bf27}, 160 (1978);
H.~J. Carmichael, J. Phys. B {\bf13}, 3551 (1980).


\bibitem{Holstein40}
T. Holstein and H. Primakoff, Phys. Rev. {\bf58}, 1098 (1940).


\bibitem{Morrison07}
S. Morrison and A.~S. Parkins, arXiv:0711.2325


\bibitem{EntanglementCriteria}
J.~K. Korbicz, J.~I. Cirac, and M. Lewenstein, Phys. Rev. Lett. {\bf95}, 120502 (2005);
{\it ibid}. {\bf95}, 259901 (2005).


\bibitem{maxC_varphi}
Note also that near $h_+^{\rm c}$ ($h_-^{\rm c}$) the entanglement $C_\varphi$ is maximized around $\varphi\simeq0$ ($\varphi\simeq\pi/2$).


\bibitem{Schneider02}
S. Schneider and G.~J. Milburn, Phys. Rev. A {\bf65}, 042107 (2002).


\bibitem{vonCube06}
Ch. von Cube {\em et al}., Fortschr. Phys. {\bf54}, 726 (2006); J. Klinner et. al, Phys. Rev. Lett. {\bf 96}, 023002 (2006)



\end{thebibliography}
\end{document}